\documentclass[aps, prb, superscriptaddress, secnumarabic, twocolumn, amssymb, longbibliography, nobibnotes]{revtex4-2}

\usepackage{amsmath,mathtools} 
\usepackage[normalem]{ulem}
\usepackage{graphics,color}

\begin{document}


\title{Floquet Graphene Antidot Lattices}

\author{Andrew Cupo}
\email{andrew.cupo@dartmouth.edu}
\affiliation{Department of Physics and Astronomy, Dartmouth College, Hanover, New Hampshire, 03755, USA}

\author{Emilio Cobanera}
\affiliation{Department of Physics, SUNY Polytechnic Institute, Utica, New York, 13502, USA}
\affiliation{Department of Physics and Astronomy, Dartmouth College, Hanover, New Hampshire, 03755, USA}

\author{James D. Whitfield}
\affiliation{Department of Physics and Astronomy, Dartmouth College, Hanover, New Hampshire, 03755, USA}

\author{Chandrasekhar Ramanathan}
\affiliation{Department of Physics and Astronomy, Dartmouth College, Hanover, New Hampshire, 03755, USA}

\author{Lorenza Viola}
\email{lorenza.viola@dartmouth.edu}
\affiliation{Department of Physics and Astronomy, Dartmouth College, Hanover, New Hampshire, 03755, USA}


\begin{abstract}
We establish the theoretical foundation of the Floquet graphene antidot lattice, whereby massless Dirac fermions are driven periodically by a circularly polarized electromagnetic field, while having their motion excluded from an array of nanoholes. The properties of interest are encoded in the quasienergy spectra, which are computed non-perturbatively within the Floquet formalism. We find that a rich Floquet phase diagram emerges as the amplitude of the drive field is varied. Notably, the Dirac dispersion can be restored in real time relative to the gapped equilibrium state, which may enable the creation of an optoelectronic switch or a dynamically tunable electronic waveguide. As the amplitude is increased, the ability to shift the quasienergy gap between high-symmetry points can change which crystal momenta dominate in the scattering processes relevant to electronic transport and optical emission. Furthermore, the bands can be flattened near the $\Gamma$ point, which is indicative of selective dynamical localization. Lastly, quadratic and linear dispersions emerge in orthogonal directions at the $M$ point, signaling a Floquet semi-Dirac material. Importantly, all our predictions are valid for experimentally accessible near-IR radiation, which corresponds to the above bandwidth limit for the graphene antidot lattice. Cycling between engineered Floquet electronic phases may play a key role in the development of next-generation on-chip devices for optoelectronic applications.
\end{abstract}


\maketitle


\section{Introduction}

Ever since graphene was fully isolated and characterized in 2004, it has remained the platform of choice for investigating massless Dirac fermions in two space dimensions (2D) \cite{neto2009electronic}. Graphene is a strong 2D material that can be easily and stably shaped into, for example, nanoribbons. The interplay between confinement and quantum effects (in the following, quantum confinement for short) opens an electronic band gap that increases with decreasing width \cite{han2007energy}. Other experimentally accessible graphene nanostructures, and the focus of this paper, are the antidot lattices of Ref.\,\onlinecite{pedersen2008graphene}. These structures are fabricated by patterning a periodic array of nanoholes into a graphene monolayer. Comparable to varying the width of a nanoribbon, varying the size of the supercell and/or the diameter of the holes of the graphene antidot lattice amounts to controlling quantum confinement and the electronic band gap. Subsequent work has showed that the shape of the holes, edge termination, and edge magnetism can have a profound effect on the way that the energy gap scales with the geometric parameters of the antidots \cite{pedersen2008optical, vanevic2009character, furst2009density, petersen2009quasiparticle, furst2009electronic, liu2009band, martinazzo2010symmetry, ouyang2010electronic, petersen2011clar, zhang2011band, ouyang2011bandgap, petersen2012clar, pedersen2012transport, liu2013universal, ouyang2014antidot, trolle2013large, brun2014electronic, thomsen2014dirac, ouyang2014bandgap, tang2014improved}. These structures can be fabricated by lithographic methods \cite{bai2010graphene, jessen2019lithographic}.

In addition to geometric control, the electronic properties of a material can be modified by time-periodic driving via an electromagnetic field. As coherent-control capabilities have continued to advance in different experimental platforms, the physics of periodically driven ``Floquet quantum matter'' has attracted growing attention across quantum science. While the subject has been 
reviewed extensively elsewhere \cite{bukov2015universal, basov2017towards, oka2019floquet, de2019floquet, harper2020topology, rudner2020band, rodriguez2021low}, here we only highlight a few results for graphene. One can open and control gaps in the quasienergy spectrum by adjusting the polarization, photon energy, and electric field amplitude \cite{syzranov2008effect, lopez2008analytic, oka2009photovoltaic, kibis2010metal, calvo2011tuning, zhou2011optical, savel2011massless, calvo2012laser, calvo2013non}. It is also possible to induce non-trivial topology in ``bulk" graphene with time-reversal symmetry-breaking circularly polarized light. As a consequence, chiral edge states appear for the corresponding system with open boundaries \cite{gu2011floquet, perez2014floquet, zhai2014photoinduced, usaj2014irradiated, perez2015hierarchy, puviani2017dynamics, wang2021floquet}, which are predicted to be detectable spectroscopically \cite{sentef2015theory, schuler2020circular, chen2020observing}. The dynamical generation of non-trivial gapped topology explains the observation of the Floquet anomalous quantum Hall effect in graphene \cite{mciver2020light, sato2019microscopic}.

In this paper we investigate the interplay between the spatial complexity of the graphene antidot lattice and a periodic driving force, and show that it leads to Floquet electronic phases beyond what is possible with standard graphene. A main motivation for our analysis stems from the fact that, for standard graphene, the above bandwidth limit can only be reached with extreme UV radiation, which is ionizing and not practical to produce at high intensities. By contrast, for the graphene antidot lattice we find that the same limit corresponds to experimentally realizable infrared (IR) photon energies. Moreover, while nanoribbons and quantum dots are also viable platforms for generating geometrically complex electronic states, the antidot lattice is the only one that is spatially extended in 2D. 

We follow a well established approach for modeling the electronic structure of the graphene antidot lattices. Our starting point is the 2D massless Dirac Hamiltonian (Sec.\,\ref{sec:GAL}) to which we add a circularly polarized electromagnetic field within the minimal coupling scheme. The properties of the periodically driven system are characterized by quasienergy spectra, computed from the extended space formulation of the Floquet formalism (Sec.\,\ref{sec:FGAL}). By varying the electric field amplitude on a fine grid for a fixed photon energy, we identify several parameter regimes where the quasienergy bands take on interesting characteristics (Sec.\,\ref{sec:QEB}). Building on these characteristics, we identify three potential quantum engineering applications: a low-wavelength-pass electronic filter (Sec.\,\ref{sec:QEB}), an optoelectronic switch, and a dynamically tunable electronic waveguide (Sec.\,\ref{sec:app}). We conclude in Sec.\,\ref{sec:SO} with a summary and outlook. 

 
\section{Background: Graphene Antidot Lattices}
\label{sec:GAL}

A graphene antidot lattice is the system that results from etching a periodic array of holes (``antidots") into a sheet of monolayer graphene. The holes should not be so large or packed so close together that they compromise the integrity of the free standing sheet. In this paper we will rely on the effective model of the graphene antidot lattices introduced and investigated in Refs.\,\onlinecite{pedersen2008graphene, furst2009electronic, brun2014electronic}. The starting line is the effective 2D massless Dirac fermion Hamiltonian for graphene,
\begin{equation}
H_{\textrm{Dirac}} = v_F \boldsymbol p \cdot \boldsymbol \sigma.
\label{ham_start}
\end{equation}
Here $v_F$ is the Fermi velocity, $\boldsymbol p$ is the momentum operator, and $\boldsymbol \sigma=[\sigma_x,\sigma_y]$ is the vector of \(x\) and \(y\) Pauli spin matrices. The internal degree of freedom is due to pseudo-spin, not the physical spin angular momentum degree of freedom, which is not included in this description. The Fermi velocity is calculated to be $v_F = 3 \tau d/(2\hbar)$, where $\tau = 2.7$ eV is the nearest-neighbor hopping parameter and $d = 0.142$ nm is the carbon-carbon bond length. We note that our starting point for the analysis, Eq.\,\eqref{ham_start}, only describes the $K$ point of graphene. To capture intervalley scattering, one should also simultaneously include the $K'$ point \cite{rodrigues2016intervalley}, which is accomplished by making the substitution $\boldsymbol \sigma \rightarrow \boldsymbol \sigma^*$ in Eq.\,\eqref{ham_start} \cite{neto2009electronic}. This can be considered as a next step in future work.

To model the confinement of massless Dirac fermions in a nanostructure, a ``mass term" $\Delta(\boldsymbol r)$ is added to Eq.\,\eqref{ham_start}: 
\begin{equation}
H_{0} = v_F \boldsymbol p \cdot \boldsymbol \sigma
+
\Delta(\boldsymbol r) \sigma_z
\label{ham_mass},
\end{equation}
where the function $\Delta(\boldsymbol r)$
is defined at each point $\boldsymbol r$ in space and takes the value zero inside and $\Delta_0$ outside of the material. The goal is to simulate a ``hard wall" barrier in the limit where \(\Delta_0\rightarrow \infty\). In practice, setting $\Delta_0 = 170$ eV suffices \cite{brun2014electronic}. 

A previous study \cite{brun2014electronic} compared the simple, continuum Dirac Hamiltonian approach outlined here to a more refined description based on a tight-binding model which, in particular, can account for the distinct edge terminations (armchair, zigzag, etc.) of the graphene lattice. The authors found that minimizing the lengths of the zigzag edges around the holes reduces the presence of localized edge states, improving quantitative agreement between the two models. From a different perspective, one expects that edge effects will play a minor role in the overall properties of the system provided that dangling bonds are hydrogen passivated and edge spins are scrambled at ambient temperature. These points justify using a continuum Dirac Hamiltonian approach.

In this work, we focus on a representative triangular antidot lattice with circular (Fig.\,\ref{gal_cir}a) or triangular holes (Fig.\,\ref{gal_tri}a). The time-independent Schr\"{o}dinger equation is
\begin{equation}
H_{0} \varphi_{n \boldsymbol k}(\boldsymbol r) = E_{n \boldsymbol k} \varphi_{n \boldsymbol k}(\boldsymbol r),
\label{TISE}
\end{equation}
where $n$ is the band index and $\hbar \boldsymbol k$ is the crystal momentum. Its solutions can be written in the Bloch form,
\begin{equation}
\varphi_{n \boldsymbol k}(\boldsymbol r) = e^{i \boldsymbol k \cdot \boldsymbol r} u_{n \boldsymbol k}(\boldsymbol r).
\label{varphi_nk}
\end{equation} 
Thanks to its periodicity in space, $u_{n \boldsymbol k}(\boldsymbol r)$ can be written as a Fourier series
\begin{equation}
u_{n \boldsymbol k}(\boldsymbol r) = \sum_{\boldsymbol G}^{} u_{n \boldsymbol k}^{(\boldsymbol G)} e^{i \boldsymbol G \cdot \boldsymbol r},
\label{u_nk}
\end{equation}
where $\boldsymbol G = a \boldsymbol G_1 + b \boldsymbol G_2$, $\boldsymbol G_1$ and $\boldsymbol G_2$ are the supercell reciprocal lattice vectors, and $a$ and $b$ are integers. The equation to solve is then
\begin{equation}
\sum_{\boldsymbol G'}^{} \mathcal{H}_{\boldsymbol k}^{(\boldsymbol G,\boldsymbol G')} u_{n \boldsymbol k}^{(\boldsymbol G')} = E_{n \boldsymbol k} u_{n \boldsymbol k}^{(\boldsymbol G)},
\label{H_G_Gprime_eig}
\end{equation}
where
\begin{equation}
\mathcal{H}_{\boldsymbol k}^{(\boldsymbol G,\boldsymbol G')} = 
\begin{bmatrix}
\Delta_{\boldsymbol G - \boldsymbol G'} & T_{\boldsymbol k \boldsymbol G} \delta_{\boldsymbol G,\boldsymbol G'} \\ 
T_{\boldsymbol k \boldsymbol G}^* \delta_{\boldsymbol G,\boldsymbol G'} & -\Delta_{\boldsymbol G - \boldsymbol G'}
\end{bmatrix} ,
\label{H_G_Gprime}
\end{equation}
with
\begin{equation}
\Delta_{\boldsymbol G - \boldsymbol G'} \equiv \frac{1}{A_\textrm{SC}} \int_{}^{} d^2r \,\Delta(\boldsymbol r) e^{-i (\boldsymbol G - \boldsymbol G') \cdot \boldsymbol r},
\label{Delta_G_Gprime}
\end{equation}
and
\begin{equation}
T_{\boldsymbol k \boldsymbol G} \equiv \hbar v_F [(k_x + G_x) - i(k_y + G_y)].
\end{equation}
The integration in Eq.\,\eqref{Delta_G_Gprime} is over the supercell of area $A_\textrm{SC}$. 
In principle, Eq.\,\eqref{H_G_Gprime_eig} is an infinite-dimensional matrix equation for each fixed \(\boldsymbol k\) and \(n\).
In practice, by forming a block matrix $\mathcal{H}_{\boldsymbol k}$ using all combinations of $\boldsymbol G$ and $\boldsymbol G'$ such that $a,b,a',b' \in [-N_{\textrm{rec}}, N_{\textrm{rec}}]$ in 
Eq.\,\eqref{H_G_Gprime}, 
Eq.\,\eqref{H_G_Gprime_eig} can be rewritten as a finite matrix diagonalization problem:
\begin{equation}
\mathcal{H}_{\boldsymbol k} \boldsymbol u_{n \boldsymbol k} = E_{n \boldsymbol k} \boldsymbol u_{n \boldsymbol k}, 
\end{equation} 
where the components of $\boldsymbol u_{n \boldsymbol k}$ contain all of the $u_{n \boldsymbol k}^{(\boldsymbol G)}$ in the allowed range. They are then used to construct approximate $u_{n \boldsymbol k}(\boldsymbol r)$ (Eq.\,\eqref{u_nk}) and, finally, the $\varphi_{n \boldsymbol k}(\boldsymbol r)$ (Eq.\,\eqref{varphi_nk}). 
For numerical evaluation in the model system under study, truncation at $N_{\textrm{rec}} = 16$ spatial Fourier modes converges the electronic band structures $E_{n \boldsymbol k}$.


\section{Floquet Graphene Antidot Lattices}
\label{sec:FGAL}

To investigate the effect of the applied control field, we rely on the standard prescription for minimally coupling the electromagnetic field to the Dirac Hamiltonian of Eq.\,\eqref{ham_mass}. The result is the time-dependent Hamiltonian
\begin{equation}
H(t) = H_0(\boldsymbol p \rightarrow \boldsymbol p + |e| \boldsymbol A(t)) = H_0 + |e| v_F \boldsymbol A(t) \cdot \boldsymbol \sigma, 
\label{ham_time}
\end{equation} 
where the vector potential $\boldsymbol A$ corresponds to a homogeneous, in-plane electric field $\boldsymbol E = -\partial_t \boldsymbol A$ with circular polarization, that is,
\begin{equation}
\boldsymbol A(t) = \frac{E_0}{\Omega} [\textrm{cos}(\Omega t),\textrm{sin}(\Omega t),0].
\label{circular}
\end{equation} 
Thus, the control parameters are the electric field amplitude $E_0$ and the angular frequency $\Omega$. The vector potential of Eq.\,\eqref{circular} misrepresents the magnetic flux density ($\boldsymbol B = \nabla \times \boldsymbol A = \boldsymbol 0$). Nonetheless, it is the typical starting point for many other investigations of radiation driven graphene and we will follow this practice \cite{syzranov2008effect, lopez2008analytic, oka2009photovoltaic, kibis2010metal, calvo2011tuning, zhou2011optical, savel2011massless, calvo2012laser, calvo2013non, gu2011floquet, perez2014floquet, zhai2014photoinduced, usaj2014irradiated, perez2015hierarchy, puviani2017dynamics, wang2021floquet, sentef2015theory, schuler2020circular, chen2020observing, topp2019topological, li2020floquet, vogl2020floquet, vogl2020effective, lovey2016floquet, delplace2013merging, liu2019engineering}. One can roughly assess the impact of ignoring the magnetic field by noting that the ratio of the magnitude of the magnetic to the electric force is at worst $v_F/c \approx 0.003$. In addition, the $\Omega t - \boldsymbol K \cdot \boldsymbol r$ that would normally appear
as the argument of the cosine and sine functions in Eq.\,\eqref{circular} reduces to $\Omega t$ since the relevant wave-vector can be chosen to be $\boldsymbol K = [0,0,K_z]$ and the graphene layer can be placed in the $z = 0$ plane.

The next step is to move to the matrix representation of the time-dependent Hamiltonian $H(t)$ (Eq.\,\eqref{ham_time}), in the basis obtained from solving the time-independent problem, recall Eq.\,\eqref{TISE}. The matrix elements are
\begin{equation}
H_{nn';\boldsymbol k}(t) = 
E_{n \boldsymbol k} \delta_{nn'} + 
|e| v_F \boldsymbol A(t) \cdot \int_{}^{} d^2r \varphi^\dagger_{n \boldsymbol k}(\boldsymbol r) \boldsymbol \sigma \varphi_{n' \boldsymbol k}(\boldsymbol r).
\label{matrix_elements}
\end{equation}
The driving renormalizes the individual bands and also generates inter-band coupling. For simplicity we focus on the two band model that emerges from keeping only the first band below and first band above the Fermi energy (0 eV). From here forward, $H_{\boldsymbol k}(t)$ refers to the time-dependent Hamiltonian in its matrix representation.

Since $H_{\boldsymbol k}(t)$ is periodic in time with period $T = 2 \pi/\Omega$, the solution of the time-dependent Schr\"{o}dinger equation, 
\begin{equation}
i \hbar \partial_t \psi_{n \boldsymbol k}(t) = H_{\boldsymbol k}(t) \psi_{n \boldsymbol k}(t),
\label{TDSE}
\end{equation} 
can be constructed using the extended space formulation of the Floquet formalism. Then the above time-dependent problem is formally mapped to a time-independent problem of diagonalizing an associated Hermitian operator defined on an enlarged space (compared to the original, physical Hilbert space) \cite{rudner2020floquet}. The solutions take the form
\begin{equation}
\psi_{n \boldsymbol k}(t) = e^{-i \epsilon_{n \boldsymbol k} t/\hbar} \Phi_{n \boldsymbol k}(t), \quad
\Phi_{n \boldsymbol k}(t+T) = \Phi_{n \boldsymbol k}(t).
\label{psi_nk}
\end{equation}
The periodicity of $H_{\boldsymbol k}(t)$ and $\Phi_{n \boldsymbol k}(t)$ allows for the Fourier decompositions in terms of time harmonics, 
\begin{equation}
H_{\boldsymbol k}(t) = \sum_{m}^{} H_{\boldsymbol k}^{(m)} e^{-i m \Omega t},
\end{equation}
\begin{equation}
\Phi_{n \boldsymbol k}(t) = \sum_{m}^{} \phi_{n \boldsymbol k}^{(m)} e^{-i m \Omega t},
\end{equation}
where $m\in {\mathbb Z}$ is the temporal Fourier index. Eq.\,\eqref{TDSE} can then be rewritten as
\begin{equation}
\sum_{m'}^{} 
\tilde H_{\boldsymbol k}^{(m,m')}
\phi_{n \boldsymbol k}^{(m')} 
= 
\epsilon_{n \boldsymbol k} 
\phi_{n \boldsymbol k}^{(m)},
\label{H_m_mprime_eig}
\end{equation}
in terms of 
\begin{equation}
\begin{split}
\tilde H_{\boldsymbol k}^{(m,m')}
=
\frac{1}{T} \int_{0}^{T} dt H_{\boldsymbol k}(t) e^{i (m-m') \Omega t}
-
\delta_{mm'} m \hbar \Omega \boldsymbol 1
\\
=
H_{\boldsymbol k}^{(m-m')}
-
\delta_{mm'} m \hbar \Omega \boldsymbol 1.
\end{split}
\label{H_m_mprime}
\end{equation}
Due to the simple form of the vector potential of Eq.\,\eqref{circular}, only matrix blocks with $|m-m'| \leq 1$ can be non-zero. By forming a block matrix $H_{\boldsymbol k}^{\textrm{(Floquet)}}$ using all values of $m,m'$ between $-N_{\textrm{Floquet}}$ and $N_{\textrm{Floquet}}$ in Eq.\,\eqref{H_m_mprime}, Eq.\,\eqref{H_m_mprime_eig} can then be rewritten as a matrix diagonalization problem
\begin{equation}
H_{\boldsymbol k}^{\textrm{(Floquet)}}
\boldsymbol \phi_{n \boldsymbol k}
=
\epsilon_{n \boldsymbol k} 
\boldsymbol \phi_{n \boldsymbol k}.
\end{equation}
For all cases to be considered below, truncation at $N_{\textrm{Floquet}} = 5$ ensures converged quasienergy spectra $\epsilon_{n \boldsymbol k}$.


\section{Floquet Band Engineering in Graphene Antidot Lattices}
\label{sec:QEB}

\begin{figure*}[t]
\includegraphics[width=0.85\textwidth]{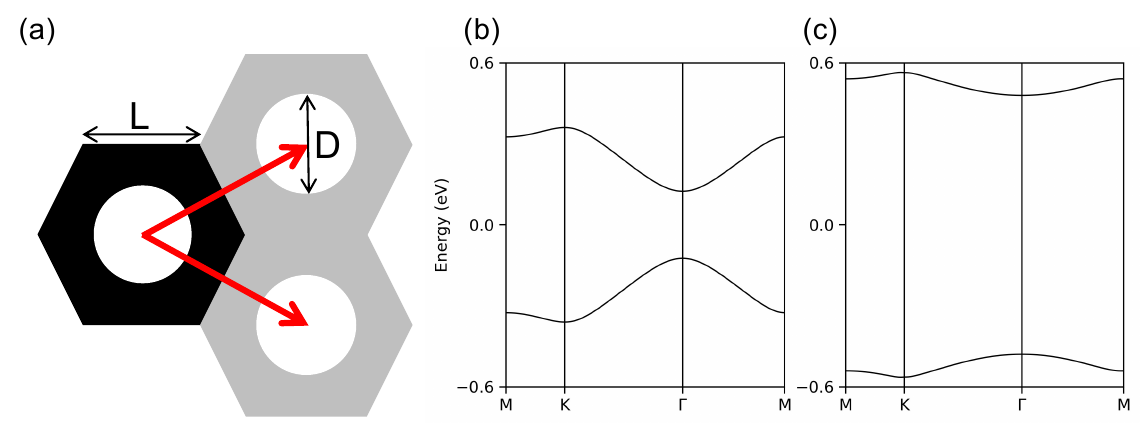}
\caption{(Color online) 
Triangular graphene antidot lattices with circular holes. (a) Geometric schematic with the superlattice vectors shown in red. The carbon lattice is not shown since we work within a continuum approach. Electronic band structure for $L = 3.0$ nm with (b) $D = 1.6$ nm and (c) $D = 3.7$ nm. The Fermi energy is located at 0 eV.}
\label{gal_cir}
\end{figure*}
\begin{figure*}[t]
\includegraphics[width=0.85\textwidth]{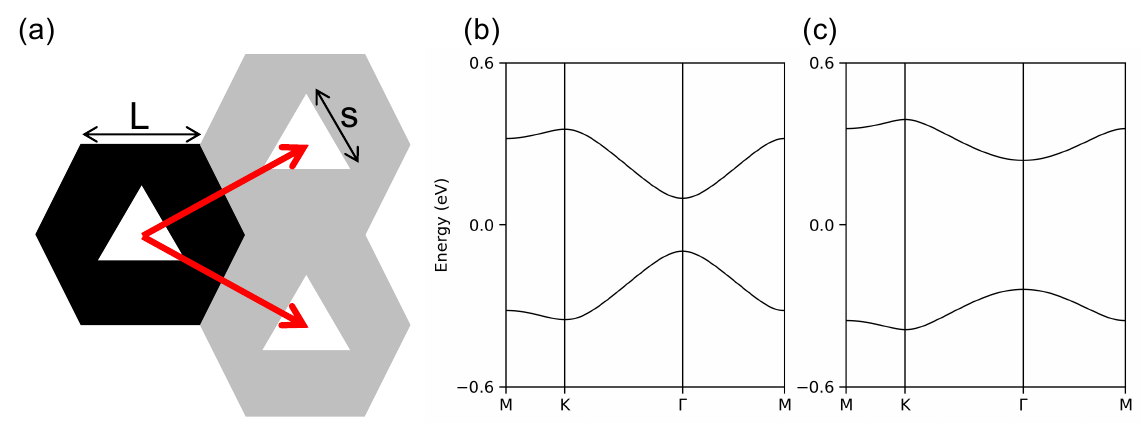}
\caption{(Color online) 
Triangular graphene antidot lattices with triangular holes. (a) Geometric schematic with the superlattice vectors shown in red. The carbon lattice is not shown since we work within a continuum approach. Electronic band structure for $L = 3.0$ nm with (b) $s = 1.4$ nm and (c) $s = 3.2$ nm. The Fermi energy is located at 0 eV.}
\label{gal_tri}
\end{figure*}

\subsection{Static system}

As a point of reference, we first calculate the electronic band structures $E_{n \boldsymbol k}$ (recall Eq.\,\eqref{TISE}), prior to irradiation, of several graphene antidot lattices. We find that, regardless of the shape of the hole, increasing the hole size while keeping the size of the supercell fixed changes the band structure from the characteristic gapless Dirac dispersion of the empty lattice (no holes) into flat bands for holes that are close to touching. Specifically, we choose the geometric parameters \(L\) and \(D\), see Fig.\,\ref{gal_cir}a, or \(L\) and \(s\), see Fig.\,\ref{gal_tri}a, so that they are consistent with the physical characteristics of graphene. The parameter \(L\) should only be on the order of a few nanometers for quantum confinement effects to play a role. In particular, we investigate circular holes of diameter \(D = 1.6\) nm and \(D = 3.7\) nm, and equilateral triangular holes of side lengths \(s = 1.4\) nm and \(s = 3.2\) nm. In all cases, \(L = 3.0\) nm. The corresponding band structures are plotted in Figs.\,\ref{gal_cir}b, \ref{gal_cir}c, \ref{gal_tri}b, and \ref{gal_tri}c. Quantum confinement opens a large band gap and adds curvature to the bands at the $\Gamma$ point of the Brillouin zone. We note that the continuum model remains valid on these length scales \cite{brun2014electronic} and that atomically precise graphene antidot lattices are within the reach of current fabrication capabilities \cite{moreno2018bottom}.

\subsection{Floquet-driven system}

With the geometric parameters specified and the time-independent problem solved, we are now well positioned to describe the periodically driven system. We choose the value of the photon energy $\hbar \Omega$ based on two criteria:

(i) The two static energy bands should automatically fall in the first Floquet Brillouin zone, that is, between $\pm \hbar \Omega / 2$. Under such a condition, there is a direct mapping between the zero electric field amplitude limit of the quasienergy spectrum $\epsilon_{n \boldsymbol k}$ and the electronic band structure $E_{n \boldsymbol k}$, as confirmed by our numerical simulations. 

(ii) In order to justify the restriction to a two band model, the photon energy should not be high enough to trigger transitions from the first valence band to the second conduction band (not shown in our figures). For the structures described in Figs.\,\ref{gal_cir} and \ref{gal_tri}, the appropriate photon energy ranges \vspace*{1mm}are 
\begin{itemize}
\item 0.72-0.81 eV \vspace*{-1mm}(Fig.\,\ref{gal_cir}b),
\item 1.13-1.15 eV \vspace*{-1mm}(Fig.\,\ref{gal_cir}c),
\item 0.71-0.77 eV \vspace*{-1mm}(Fig.\,\ref{gal_tri}b),
\item 0.78-0.85 eV (Fig.\,\ref{gal_tri}c). 
\end{itemize}
To ensure that our results are stable against small variations of the parameters, we calculate the quasienergy spectra for two values of $\hbar \Omega$ in these ranges of validity. The specific values of the photon energy are 
\begin{itemize}
    \item 0.75, 0.80 eV \vspace*{-1mm}(Fig.\,\ref{gal_cir}b),
    \item 1.13, 1.15 eV \vspace*{-1mm}(Fig.\,\ref{gal_cir}c),
    \item 0.71, 0.77 eV \vspace*{-1mm}(Fig.\,\ref{gal_tri}b),
    \item 0.78, 0.85 eV (Fig.\,\ref{gal_tri}c).
\end{itemize}

\begin{figure}[t]
\centering{\includegraphics[scale=0.9]{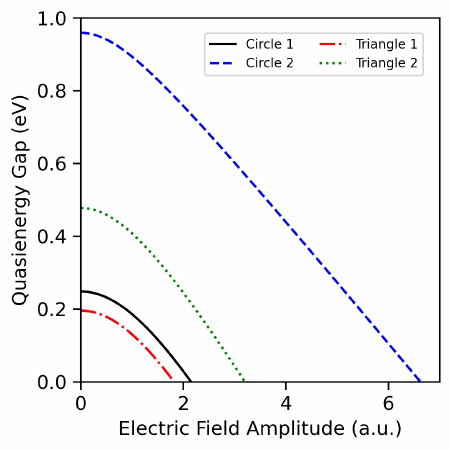}}
\caption{(Color online)
Continuous tuning of the Floquet quasienergy gap (at $\Gamma$) with the electric field amplitude. Each curve starts at the equilibrium state (0 a.u.) and ends at the Floquet Dirac phase (horizontal axis intersection). Circle 1, Circle 2, Triangle 1, and Triangle 2 correspond to the structures defined in Figs.\,\ref{gal_cir}b, \ref{gal_cir}c, \ref{gal_tri}b, and \ref{gal_tri}c.}
\label{scaling}
\end{figure}

In addition to the frequency of the radiation, one can also control the intensity. We scanned the electric field amplitude $E_0$ from 0 to 6 a.u. (1 a.u. = 1.291 (GeV/nm)/nC) in increments of 0.05 a.u., expanding and interpolating further when necessary. The units of $E_0$ are taken from a previous paper on Floquet Dirac materials \cite{hubener2017creating}. The values of $E_0$ that we investigate are all well below the pulsed damage threshold for graphene, which is 23 a.u. \cite{roberts2011response}. This said, we find that the qualitative features of the quasienergy bands are the same for all cases after a rescaling of the electric field amplitude. As a quantitative illustration of this fact, we show in Fig.\,\ref{scaling} the dependence of the quasienergy gap upon
the electric field amplitude for our four Floquet graphene antidot lattices. Based on these results, for 
succinctness we only explicitly show and discuss the results for circular holes with $D = 1.6$ nm and $\hbar \Omega = 0.8$ eV in what follows.

The insensitivity of the quasienergy spectra to the shape of the holes was an unexpected outcome of our simulations. A possible explanation is that the effective Dirac Hamiltonian only captures quantum confinement effects while softening or removing altogether the effect of particular edge configurations. The antidot lattice imposes complicated boundary conditions on the wave functions, see Eq.\,\eqref{TISE}. This added layer of richness emerges in the quasienergy spectra via the matrix elements in Eq.\,\eqref{matrix_elements}. The particle-hole symmetry of the starting Dirac Hamiltonian (Eq.\,\eqref{ham_start}) is preserved in all cases.

\begin{figure*}[t]
\includegraphics[width=0.94\textwidth]{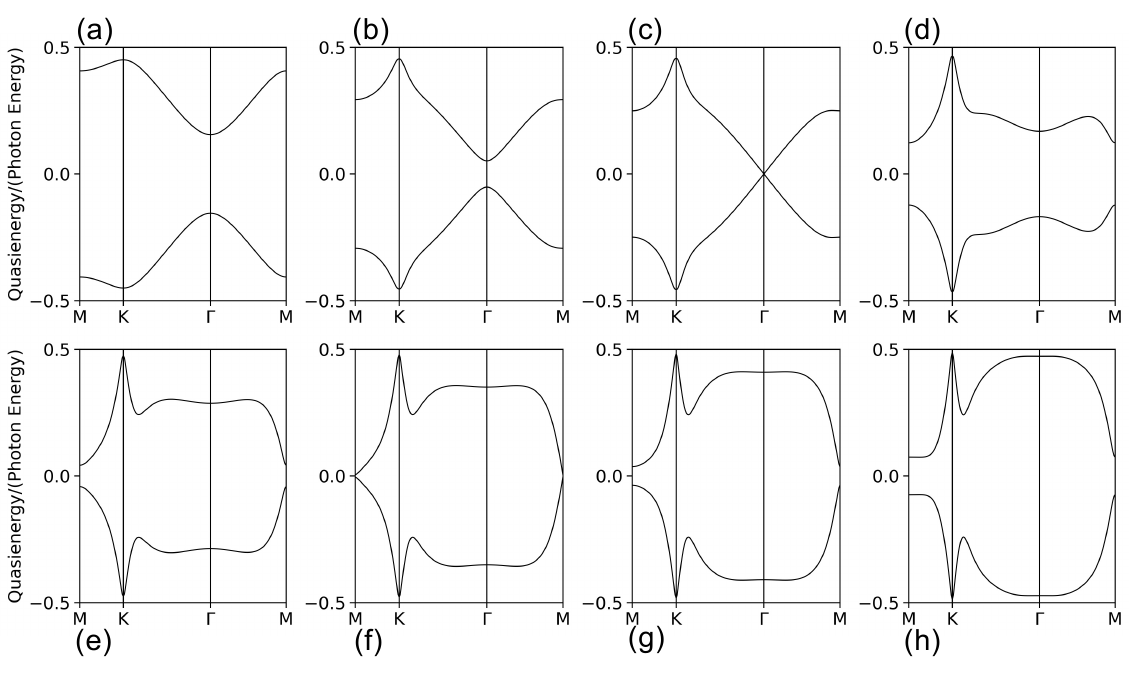}
\vspace*{-4mm}
\caption{Representative quasienergy spectra for the graphene antidot lattice defined in Fig.\,\ref{gal_cir}b, irradiated with circularly polarized light at a photon energy of 0.8 eV. $E_0$ is the electric field amplitude, where 1 a.u. = 1.291 (GeV/nm)/nC \cite{hubener2017creating}. (a) $E_0 = 0$ a.u. Zero electric field amplitude limit for reference. (b) $E_0 = 1.7$ a.u. (c) $E_0 = E_{0c_1} = 2.15$ a.u. The Dirac dispersion of standard graphene is effectively restored. (d) $E_0 = 3.4$ a.u. The quasienergy gap is shifted to the $M$ point. (e) $E_0 = 4.2$ a.u. The quasienergy bands are selectively flattened in reciprocal space near the $\Gamma$ point. (f) $E_0 =  E_{0c_2} = 4.62$ a.u. In moving from $K$ to $M$ the bands are quadratic, whereas in moving from $\Gamma$ to $M$ they are linear, indicative of a Floquet semi-Dirac material. (g) $E_0 = 5$ a.u. (h) $E_0 = 5.4$ a.u.}
\label{qe}
\end{figure*}

The quasienergy bands depicted in Fig.\,\ref{qe} summarize our main findings about the Floquet phases that the system can access as the electric field amplitude is varied in the parameter regime under exploration.  Recall that, in the absence of driving, the Dirac point of pristine graphene is replaced by parabolic bands separated by a gap at the \(\Gamma\) point for the graphene antidot lattice (see Fig.\,\ref{qe}a). Upon adding driving, a number of distinctive physical features emerge:
\begin{enumerate}
    \item The Dirac point can be {\em restored dynamically} for a suitable value of the electric field amplitude that we call \(E_{0c_1}\). The quasienergy gap closes at the \(\Gamma\) point of the Brillouin zone.
    \item 
    For \(E_0>E_{0c_1}\), the quasienergy gap reopens and moves to the \(M\) point of the Brillouin zone; the bands flatten substantially close to the \(\Gamma\) point. 
    \item There is a second value of the intensity \(E_{0c_2}>E_{0c_1}\) for which the quasienergy gap closes again but now at the \(M\) point. Moreover, the bands touch in a most peculiar way: they feature \emph{linear} dispersion in one direction and \emph{quadratic} in another.
    \item The quasienergy gap reopens for \(E_0>E_{0c_2}\).
\end{enumerate}

Thinking of the dynamical restoration of the Dirac point in the Floquet graphene antidot lattice as a phase transition, we shall call the phase with \(0\leq E_0<E_{0c_1}\) the \emph{Floquet quasi-equilibrium phase} and the phase with \(E_{0c_1}<E_0<E_{0c_2}\), which ends with the closing of the quasienergy gap at \(M\), the first Floquet phase, or \emph{Floquet Dirac phase}. There is a second Floquet phase, or \emph{Floquet semi-Dirac phase}, for \(E_0>E_{0c_2}\). It is worth it to further expand on these remarkable features and speculate on how they physically come about.

\smallskip

\underline{1.}: 
The ability to restore the Dirac point (Fig.\,\ref{qe}c) 
seems only possible because the ``mass term" in the Hamiltonian of Eq.\,\eqref{ham_mass} hides the Dirac dispersion of standard graphene, and is then dynamically canceled by the radiation field. While pristine graphene already features a Dirac point, having the non-driven gapped bands and the Floquet Dirac point enables at least two novel applications, as we argue in Sec.\,\ref{sec:app}. Furthermore, varying the geometric parameters of static graphene antidot lattices only results in discrete tunability of the electronic band gap. By contrast, the electric field amplitude can be varied continuously, which leads to an {\em adiabatic connection} between the static state and the Floquet Dirac regime. In Fig.\,\ref{scaling} we explicitly see how the quasienergy gap, which remains at $\Gamma$, can be tuned continuously and monotonically.

\smallskip

\underline{2.}: Non-ballistic transport, the phonon-assisted decays that relax excited carriers, and the lifetimes of the states involved in optical transitions all depend on the nature of the electron-phonon coupling. This, in turn, depends on the high-symmetry point where the band extrema occur \cite{giustino2017electron}. Therefore, the ability to shift the quasienergy gap from the $\Gamma$ point in the equilibrium phase to the $M$ point in the Floquet phase (Fig.\,\ref{qe}d) can affect these three fundamental processes significantly via the difference in electron-phonon coupling between high-symmetry points. These phenomena could be investigated in quantitative detail with an adapted Floquet-Boltzmann approach \cite{genske2015floquet}.

A next notable feature is localization, which manifests as band flattening. On the one hand, in the absence of driving, the bands are flattened if the holes of an antidot lattice are forced to nearly touch. This flattening is non-selective, in the sense that it occurs over the entire high-symmetry lines. On the other hand, our periodically driven system with $E_0 = 4.2$ a.u. (Fig.\,\ref{qe}e) features bands which are {\em selectively flattened} near $\Gamma$ in reciprocal space. On the basis of the Hellmann-Feynman theorem, the group velocity of the quasienergy bands is given by $\boldsymbol v_{n \boldsymbol k} = \hbar^{-1} \nabla_{\boldsymbol k} \epsilon_{n \boldsymbol k}$ \cite{schanz2005directed}. Therefore, near the $\Gamma$ point, the charge carriers have zero group velocity, which is indicative of \textit{selective} dynamical localization. This behavior persists as the electric field amplitude continues to increase, at least in the range of values considered here. Furthermore, since the band flattening occurs near $\Gamma$ and not another high-symmetry point, the system acts as a low-wavelength-pass electronic filter. For comparison, let us briefly mention here previous work \cite{agarwala2016effects}, showing that periodic kicking of standard graphene can result in \textit{complete} dynamical localization along one entire axis in reciprocal space or over the whole Brillouin zone.

\smallskip

\underline{3.}: In Fig.\,\ref{qe}f, quadratic and linear dispersions emerge from the $M$ point in orthogonal directions, which correspond to ``non-relativistic-like" and ``relativistic-like" carriers, respectively. This behavior is the signature of a Floquet semi-Dirac material, which has already been observed for standard Floquet graphene \cite{delplace2013merging, agarwala2016effects, liu2019engineering}. Even so, the physical mechanism of this striking anisotropy is not the same for all cases. In previous work, it was linearly polarized radiation \cite{delplace2013merging, liu2019engineering} or uniaxial periodic kicking \cite{agarwala2016effects} which generated the inequivalence between different directions in reciprocal space. In our setup, since we apply circularly polarized radiation, the anisotropy is a truly {\em emergent} phenomenon. In fact, in a different context, this is reminiscent of the anisotropic quasiparticle gap closing that has been found to emerge from competing interactions in both static and Floquet-driven gapless $s$-wave superconductors on 2D square lattices \cite{Deng, Poudel}.

As a platform for realizing a Floquet semi-Dirac material, Floquet graphene antidot lattices offer some distinct advantages over existing approaches. First, the semi-Dirac property can be realized at much lower photon energies ($\hbar \Omega =$ 0.72-0.81 eV), whereas previously values of around at least 4 eV were required \cite{delplace2013merging, liu2019engineering}. Second, while the work on periodic kicking is of fundamental physical interest, the proposals to realize it experimentally are not practical at the time of writing \cite{agarwala2016effects}. Third, in the case of pristine graphene, the static Dirac and the Floquet semi-Dirac points are both gapless. On the other hand, for the graphene antidot lattice, the static system is gapped and the Floquet semi-Dirac point is gapless. Having a different quasienergy gap between the static and Floquet states may enable different applications.

\smallskip

\underline{4.}: The first Floquet phase ends with the closing of the quasienergy gap at the \(M\) point. The gap reopens immediately for \(E_0>E_{0c_2}\), landing one in a second Floquet phase. We show a couple of quasienergy band structures for this phase in Figs.\,\ref{qe}g and\,\ref{qe}h. There are several other Floquet phases beyond this second one and below the damage threshold of \(E_0 = 23\) a.u. We leave them as a topic for future research.


\section{Applications}
\label{sec:app}

As we showed in the previous section, the irradiated graphene antidot lattice can cycle between gapped, quadratic bands for vanishing intensity and a dynamically induced Dirac point for a suitable value of the intensity. We recognize at least two possible applications of this behavior. In this paper we will give only a qualitative description of them, leaving a more quantitative study for the future. At this level of analysis, we envision that the operation timescale of any such Floquet device should be much slower than the drive period, so that the initial state does not change appreciably within a period and a description in terms of the Floquet stroboscopic dynamics will suffice \cite{rudner2020floquet}. In addition, and this point is specific to our setup, we do not know exactly what to expect for the mobility of the charge carriers at the dynamically induced Dirac point. We judge that a more quantitative description of our devices will have to be grounded in simulations within the Floquet Landauer-B\"{u}ttiker formalism \cite{kohler2005driven, fruchart2016probing}. 


\subsection{An optoelectronic switch}

The first application we envision is an optoelectronic switch, which could function as an optical computing element. For this device, a steady voltage drop is held across the graphene antidot lattice during operation. Hence, without driving, the gapped system should  support a small current which corresponds to the ``off" state of the switch. By contrast, with suitable driving, the gapless Floquet phase with the Dirac dispersion should allow for a much larger current corresponding to the ``on" state. Notice that the operating principle of this optoelectronic switch goes beyond the photoconductivity level (meaning that photons promote additional carriers into the conduction band). Rather, the operation mechanism is the renormalization of the band structure by the driving. Future numerical simulations should characterize the device more precisely through the on/off current ratio.


\subsection{A dynamically tunable electronic waveguide}

The second application we envision is a device that could be described as a dynamically tunable electronic waveguide. If the antidot lattice is only irradiated in a particular region of space, then one expects that the charge carriers will preferentially follow the path where the Dirac dispersion is restored. This idea presupposes that the properties we discovered still hold when translational symmetry is broken on the length scales of the waveguide. Our waveguide proposal is inspired by previous work where a graphene antidot lattice waveguide was formed by selectively leaving the graphene unpatterned, thus locally restoring the Dirac dispersion of standard graphene and creating a current channel \cite{pedersen2012graphene}. However, in that case the waveguide path was fixed, whereas in our proposal we conjecture that spatially selective irradiation should permit dynamical control of the current channels. 


\section{Conclusions and Outlook}
\label{sec:SO}

We introduce the Floquet graphene antidot lattice as a platform for investigating the properties of quantum confined and periodically driven massless Dirac fermions in 2D. The actual system is a sheet of graphene decorated by a periodic array of holes (the ``antidots") and subjected to steady irradiation. To focus on the physics of massless Dirac fermions, we model the graphene antidot lattice in terms of a Dirac Hamiltonian with a spatially varying ``mass term." Time-periodic driving is introduced by including an electromagnetic field at the minimal coupling level. We use the extended space formulation of the Floquet formalism to compute (numerically) the quasienergy band structures. 

A number of notable properties emerge for various electric field amplitudes. First, the Dirac point, which had been removed by the antidot lattice, can be dynamically restored by irradiation. Second, the quasienergy gap can be shifted from one high-symmetry point of the Brillouin zone, the \(\Gamma\) point, to another, the $M$ point, and that shift could affect dramatically the electron-phonon coupling in the system. Third, after the gap has shifted, the bands can be flattened near the $\Gamma$ point, indicative of selective dynamical localization. Finally, the quasienergy gap can again close at the $M$ point accompanied by exotic semi-Dirac behavior, where orthogonal directions emanating from the $M$ point feature quadratic and linear dispersions. Every one of these features is of intrinsic physical interest, but we have also pointed out potential device applications: an optoelectronic switch, a dynamically tunable electronic waveguide, and a low-wavelength-pass electronic filter. In general, the ability to cycle between different electronic phases with lasers may play an important role in the development of optoelectronic devices.

Some qualitative features of our quasienergy spectra can also be observed for systems different from ours, but we have made a careful case that the Floquet graphene antidot lattice may offer distinct practical advantages. From an experimental perspective, all of our predictions are valid for accessible near-IR radiation, in contrast to standard graphene, where the same non-resonant above bandwidth limit corresponds to ionizing extreme UV photon energies. Moreover, while we focused on graphene antidot lattices, everything was based on effective Dirac Hamiltonians, where a rescaling of parameters can be used to translate our results to other 2D Dirac materials. 

We conclude with some ideas for future research. One immediate question that we have not answered is whether the Floquet graphene antidot lattice features non-trivial topology \cite{zhang2015prediction, pan2017berry, kariyado2018counterpropagating}. More precisely, this system combines two gapping mechanisms, the time-reversal-symmetry-breaking circularly polarized light and the chiral-symmetry-breaking antidot lattice, both of which retain the particle-hole (charge-conjugation) symmetry. We have not determined whether the resulting quasienergy gap is topologically non-trivial, or if such properties vary based on the defined phase regimes (quasi-equilibrium, Floquet Dirac, Floquet semi-Dirac). Another interesting question is how irregularities in the placement and shapes of the holes (disorder) affect the emergent properties \cite{hung2013disorder, yuan2013electronic, power2014electronic, fan2015electronic}. And finally, there is the critical question that we have asked before: What is the mobility of charge carriers at the \textit{emergent} Dirac point? 



\section{Acknowledgements}

This work was supported by the NSF under grant No. OIA-1921199. The computations in this work were performed on the Discovery cluster and HPC environments supported by the Research Computing group, IT\&C at Dartmouth College. We thank Stephen Carr for a critical reading of this manuscript and Vincent P. Flynn for discussions about the thermal conductivity of periodically driven systems. 


\bibliography{references}


\end{document}